\documentstyle[epsfig]{mn}
\title[Geometrical estimators and CMB]{Geometrical estimators as a
test of Gaussianity in the CMB}
\author[R.B.Barreiro et al.]{R.B.Barreiro$^1$, E.Mart\'\i
nez-Gonz\'alez$^2$ and J.L.Sanz$^2$\\ 
$^1$ Astrophysics Group, Cavendish Laboratory, Madingley Road,
Cambridge CB3 0HE, UK \\
$^2$ Instituto de F\'\i sica de Cantabria, Fac. Ciencias, Avda. de los
Castros s/n, 39005 Santander, Spain \\
}
\date{Accepted ???. Received ???; in original form ???}
\begin{document}
\maketitle

\begin{abstract}
We investigate the power of geometrical estimators on
detecting non-Gaussianity in the cosmic microwave background. 
In particular the number, eccentricity and Gaussian curvature of
excursion sets above (and below) a threshold are studied.
We compare their different performance when applied to
non-Gaussian simulated maps of small
patches of the sky, which take into account the angular resolution and
instrumental noise of the Planck satellite. These non-Gaussian
simulations are obtained as perturbations of a Gaussian field in two
different ways which introduce a small level of skewness or kurtosis in
the distribution.
A comparison with a classical estimator, the
genus, is also shown. We find that the Gaussian curvature is the
best of our estimators in all the considered cases. Therefore we
propose the use of this quantity as a particularly useful test
to look for non-Gaussianity in the CMB.
\end{abstract}


\section{Introduction}
Future cosmic microwave background (CMB) experiments, such as the VSA
(Jones \& Scott 1998), the NASA MAP satellite 
(Bennett et al. 1997) and the Planck
mission from ESA (Tauber 2000), will
provide with unique data to constraint fundamental 
cosmological parameters as well as to distinguish between competing
theories of structure formation in the early universe.
%
%
In particular, the standard inflationary model predicts Gaussian
fluctuations whereas topological defects give rise to non-Gaussian
signatures in the CMB. Therefore, the study of the Gaussianity
of the CMB is a key issue to understand the nature of the primordial
perturbations that led to the formation of large-scale structure.
In addition, future high resolution maps should be searched for any
traces of non-Gaussianity since both foregrounds and systematic errors
can also leave non-Gaussian imprints.

A large number of methods have been proposed in the literature to test
the Gaussianity of the CMB (for a review see Barreiro 2000).
They include extrema correlation function (Kogut et al. 1996, Barreiro 
et al. 1998, Heavens \& Sheth 1999), properties of hot and cold spots
(Coles \& Barrow 1987, Mart\'\i nez-Gonz\'alez et al. 2000), Minkowski
functionals (Coles 1988, Gott et al. 1990, Kogut et al 
1996), bispectrum analysis (Ferreira et al. 1998, Heavens 1998,
Magueijo 2000), wavelets (Pando et al. 1998, Hobson et al 1999, 
Mukherjee et al. 2000, Barreiro et al. 2000),
multifractals (Pompilio et al. 1995) and partition function 
(Diego et al. 1999, Mart\'\i nez-Gonz\'alez et al. 2000).


In this paper we investigate the use of geometrical estimators on
detecting non-Gaussianity in the CMB using small scale simulated maps
(for Gaussian and non-Gaussian fields). These simulations take into account the
angular resolution and sensitivity predicted for two of the Planck
frequency channels (the LFI 100 GHz and the HFI 217 GHz channels).
In particular, we study the performance
of the number, eccentricity and Gaussian curvature of excursion sets
above (or below) a threshold. These quantities have been studied for a
Gaussian field by Barreiro et al. (1997). For comparison, we also
present the same analysis for the genus, a topological quantity that has
been proposed as a good estimator to test the Gaussianity of the CMB.

Simulated maps for the standard inflationary model can be very
accurately (at the level of $\simeq$ 1 per cent in the
power spectrum) obtained (e.g. Seljak \& Zaldarriaga 1996, Hu et al. 1998).
However this is not the case for models based on topological
defects. Besides, the power spectrum predicted by pure topological
defects scenarios (see e.g. Pen et al. 1997, Contaldi et al. 1999) seems
not to be in  
agreement with the latest CMB data produced by BOOMERANG (de Bernardis
et al. 2000) and MAXIMA (Hanany et al. 2000) but 
the possibility of an hybrid scenario that combines both inflation and
topological defects remains viable (Bouchet et al. 2000,
Contaldi 2000).
To take into account for all of these uncertainties, we have used 
two {\it generic} non-Gaussian fields.
In particular, physically motivated
non-Gaussian models, e.g. those based on topological defects or non-standard
inflation, generically produce certain level of skewness or/and kurtosis. Since
these are the lowest order kumulants they are also the least affected by the
presence of instrumental white noise.
Therefore, we have generated two different kind of
non-Gaussian maps obtained as weak perturbations of a Gaussian field. 
The first one, that we call $P$ distribution, has an
asymmetric 1-point density function (1-pdf) with a positive tail, what
introduces a small level of skewness. The 1-pdf of the second
non-Gaussian field, the $B$ distribution, is symmetric and have a higher peak
and longer tails than the Gaussian one. This produces
a positive value of the (excess) kurtosis.
Some of these characteristics have been shown to be produced by
topological defect models by different authors.
Turok (1996) performed sub-degree scale simulations of the CMB from cosmic
defects, finding a clear excess of hot spots for monopoles and
textures, a signature also present on our $P$
distribution. Perivolaropoulos (1993) used an analytical model to 
investigate the statistics of the CMB temperature maps induced by
topological defects. He obtained a 1-pdf for cosmic strings that
presents a central peak and long tails, resembling our $B$ distribution.

The outline of this paper is as follows. In $\S2$ we discuss the
geometrical quantities that we test as non-Gaussian estimators. $\S3$
describes our Gaussian and non-Gaussian CMB simulations.
The results, including the effect of instrumental
noise, are presented in $\S4$. Finally, the main conclusions of our
work are summarised in $\S5$.

\section{Geometrical estimators}
\begin{figure}
\begin{center}
\resizebox{5cm}{!}{\includegraphics{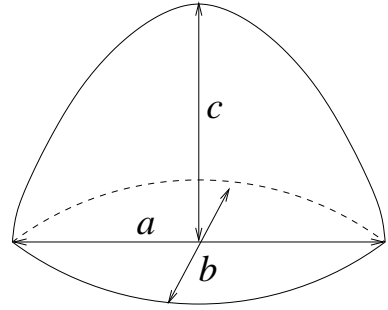}}
\end{center}
\caption[]
{To obtain the mean eccentricity and Gaussian curvature of the
excursion sets of the CMB, these are fitted to a paraboloid as the one
shown in the figure. 
$a$ and $b$ are the major and minor axes of the ellipse that
forms the basis. $c$ is the distance between the geometrical centre of
the ellipse and the maximum of the paraboloid}
\label{paraboloide}
\end{figure}

We have already pointed out the importance of testing the Gaussianity
of the CMB. An interesting possibility is
the use of geometrical estimators related
to excursion sets above or below a given threshold. In particular,
we have considered the mean number, eccentricity and Gaussian
curvature of regions above or below a threshold $\nu$ (in units of the
signal dispersion).
All these quantities can be easily estimated from a map. The
procedure is as follows. First the excursion sets, i.e. connected
pixels, above $\nu$ (and below $-\nu$) are identified. Each of them is
fitted to a paraboloid with parameters $a,b$ and $c$ (see
Fig.~\ref{paraboloide}). The eccentricity $\varepsilon$ and
Gaussian curvature $\kappa$ of each of the spots are then estimated as
\begin{eqnarray}
\varepsilon&=&\sqrt{1-\left(\frac{b}{a}\right)^2} \nonumber \\
\kappa&=&\frac{c}{ba}
\end{eqnarray}
where $a,b$ are in arcminutes and $c$ in units of the map dispersion.
Only those excursion sets with a
single maximum (or minimum when below the threshold) and formed by at
least eight pixels are 
considered to obtain the mean Gaussian curvature and eccentricity of
the map.

In addition, these quantities can be obtained (semi)analytically for a
homogeneous and isotropic Gaussian field, at least locally around the
maximum. 
On the one hand, the probability density functions of the number,
eccentricity and Gaussian curvature of maxima (which coincide
asymptotically with the excursion sets) for a homogeneous and
isotropic Gaussian field have been studied by Barreiro et al. (1997).
On the other hand, the expected number of excursion sets $<N>$ over the
whole sphere can be estimated as (Vanmarcke 1983):
\begin{equation}
<N>=\frac{2}{\pi\theta_c^2}\frac{e^{-\nu^2}}{{\rm erfc}(\nu/\sqrt2)}
\label{vanm}
\end{equation}
where $\theta_c$ is the coherence angle of the field that depends only
on the 2-point correlation function.

For comparison, we have also extended our analysis to the
genus, a topological quantity that has been commonly used to test the
Gaussianity of the CMB.
The genus at a given threshold can be approximately estimated as the
number of isolated high-temperature regions minus the number of
isolated low-temperature regions. The mean genus $<g>$ per unit area
for a homogeneous and isotropic Gaussian field is given by
\begin{equation}
<g>=\frac{1}{(2\pi)^{3/2}\theta_c^2}\nu e^{-\nu^2/2}
\label{genero}
\end{equation}
To estimate this quantity from a map we have used a method 
based on adding up the contribution to the genus from
each vertex of the pixelised map. This method is described in detail in
Jones (1999) (see also Hamilton et al., 1986).

\begin{figure}
\begin{center}
\resizebox{\hsize}{!}
{\includegraphics{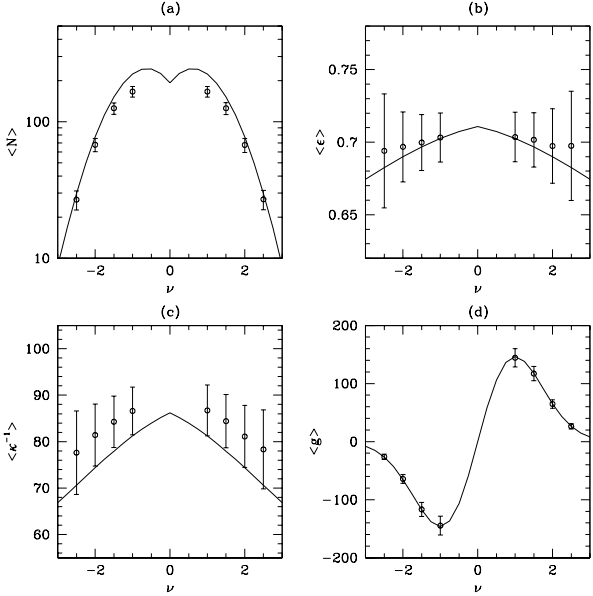}}
\end{center}
\caption[]
{Comparison of the average values of our estimators as computed from
500 simulated maps (open circles) with an approximation of the theoretical
expectations for a Gaussian field (solid lines). The error bars show
the dispersion obtained from the simulations. The panels correspond
to (a) number of regions, (b) eccentricy and (c) inverse of the Gaussian 
curvature for excursion sets above (or below for negative $\nu$) 
a threshold $\nu$, and (d) genus.}
\label{teorico}
\end{figure}

In order to study the validity of our algorithms on estimating the above
quantities, we have compared the values computed from simulations with the
theoretical expectations for the Gaussian case.
Fig.~\ref{teorico} shows the mean number of regions, eccentricty, the
inverse of the
Gaussian curvature and genus obtained from 500 Gaussian
realisations for a $\Lambda$-CDM model. The size of the maps is
512*512 pixels, with a pixel size of 1$'$.5 and they have been
smoothed with a Gaussian beam of FWHM=5.$'$5 .
The solid lines correspond to an approximation of the theoretical
expectations for each of these quantities. For the number of regions
and genus, the curves have been obtained from equations \ref{vanm} and
~\ref{genero} respectively, whereas the inverse of the
Gaussian curvature and the eccentricity were calculated using
the probability density functions given by Barreiro et al.(1997).
We can see that there is a certain difference between the computed and
predicted values for the number of regions, especially at low
thresholds. This is due to the fact
that the Vanmarcke conjecture is an approximation that overestimates
the number of excursion sets at low thresholds.
Regarding the inverse of the Gaussian curvature as well as the
eccentricity, we have to point out that the theoretical values have
been obtained locally around the maxima, whereas on the simulations 
we consider the whole
of the excursion set and discard those with more than a single
maximum and less than eight pixels. This procedure biases the expected
values and can explain the discrepancies between the
theoretical and computed values, that are however reasonably close.
Finally, as expected, the values computed from the simulations for the
genus are in good agreement with the theoretical curve.

\section{Simulations}

In order to test the performance of our estimators, we have
simulated small patches of the sky of size 12.8 square degrees with
Gaussian and non-Gaussian statistical properties.
We have considered two different angular resolutions that correspond
approximately to two
of the Planck frequency channels. On the one hand, we have produced maps
with pixel size of 3 arcminutes, smoothed with a Gaussian beam of
FWHM=10$'$, which corresponds to the resolution of the LFI 100 GHz channel. 
On the other hand, we have simulated maps with the characteristics of
the HFI 217 GHz channel, i.e., pixel size of 1.5 arcminutes and
Gaussian beam of FWHM=$5'.5$. We have also taken into account the
instrumental noise of these channels.

The Gaussian simulated maps have been generated from a CDM model
normalised to COBE/DMR and parameters
$\Omega_m=0.3$, $\Omega_\Lambda=0.7$, $h=0.5$, $\Omega_b=0.05$ and
$n=1$. The power spectrum of such a model has been obtained with
CMBfast (Seljak \& Zaldarriaga 1996).
The non-Gaussian simulations have been produced as follows. First, an
uncorrelated 2D Gaussian map with zero mean and unit variance 
is obtained. This map is then transformed
in two different ways according to the following expressions
(Weinberg \& Cole 1992):
\begin{eqnarray}
F_1&=&\frac{e^{aG}-e^{a^2/2}}{\sqrt{e^{2a^2}-e^{a^2}}}
~~~~~~~~~~~~~~~~~P~~{\rm transformation} \\
F_2&=&\frac{G(e^{aG}+e^{-aG})}{\sqrt{2\left[1+(1+4a^2)e^{2a^2}\right]}}
~~~B~~{\rm transformation}
\end{eqnarray}
where G corresponds to the initial Gaussian field and $F_1$,$F_2$ to
the transformed fields. The level of
non-Gaussianity introduced can be controlled with the parameter
$a$. In particular when $a \to 0$ we recover the Gaussian case. 
In this way, we
obtain two kind of uncorrelated non-Gaussian fields with zero mean and unit
variance. 
Next, we rescale the amplitude of each Fourier
mode to obtain the same power spectrum as for the Gaussian CDM
model.
By rescaling the power spectrum the 1-pdf is modified
with respect to the uncorrelated one, but it is still non-Gaussian
(hereinafter, when we refer to $P$ and $B$ transformations, it should be 
understood the transformations given by (4) and (5) plus a subsequent 
rescaling of the power spectrum as explained before). In
particular, for the $P$ transformation we are constructing a
non-Gaussian field whose distribution is skewed with a positive tail,
whereas for the $B$ transformation, the distribution is broadened with
both positive and negative tails. Finally, the new maps are
smoothed with the required Gaussian beam. For both kind of
non-Gaussian fields, we 
have considered small enough values of $a$ such that their 1-point
distributions are indistinguishable from the Gaussian one. In
addition, the Gaussian and non-Gaussian simulations share the same
power spectrum by construction.
\begin{figure}
\resizebox{\hsize}{!}{\includegraphics
{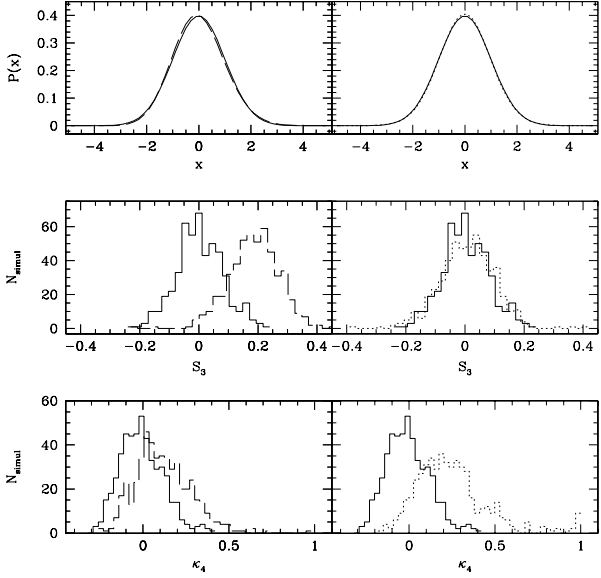}}
\caption[]
{Top: 1-pdf function for the Gaussian (solid line), $P$ (dashed line) and
$B$ (dotted line) distributions (with unit dispersion) obtained
averaging over 500 
simulations. The parameter $a=0.8$ has been used for the transformed
fields. The size of the simulated maps was 12.8 square degrees with a
pixel of $1.5$ arcminutes and they were smoothed with a Gaussian beam
with FWHM=$5.5$ arcminutes, which corresponds to the angular
resolution of the Planck 217 GHz channel. The corresponding histograms
for the skewness (middle) and kurtosis (bottom) for the same
simulations are also plotted.}
\label{1pdf}
\end{figure}
Fig.~\ref{1pdf} (top panels) shows the 1-pdf for the Gaussian and
non-Gaussian fields for $a=0.8$ that have been obtained averaging over
500 simulations with angular resolution corresponding to the 217 GHz
channel. Noise has not been included in this case. For the sake of
clarity, we have not plotted the value of the dispersion at each bin
of the histogram as error bars
but they completely overlap for the Gaussian and
non-Gaussian 1-pdf's. We can see that the $P$
distribution has a positive skewness, whereas the $B$ one
presents a small level of kurtosis.
This can also be seen in the middle and bottom panels, where the
histogram of the values of the skewness and kurtosis have been plotted
for the 500 Gaussian and non-Gaussian simulations (these quantities
have been obtained using unbiased estimators as explained in Hobson
et al. 1999).
Table~\ref{skecur} gives the average values as well as the dispersions
of the skewness and kurtosis distributions obtained from the previous
simulations.
From these values, we can see that the Gaussian and non-Gaussian
distributions can be differentiated only at the $\sim 1\sigma$ level 
using the skewness and kurtosis.
\begin{table}
\caption[]{Values of the mean and dispersion of the skewness and
kurtosis distributions obtained from 500 simulations.}
\label{skecur}
\begin{center}
\begin{tabular}{|c|c|c|c|c|}
\hline
Distribution & $<S_3>$ &$\sigma(S_3)$& $<K_4>$& $\sigma(K_4)$ \\
\hline
G & $3.72\times10^{-3}$ & 7.80$\times 10^{-2}$ & -6.55 $\times 10^{-3}$ &
0.125 \\
P ($a$=0.8) & 0.188 & 8.60 $\times 10^{-2}$ & 0.117 & 0.161 \\
B ($a$=0.8) & 7.35 $\times 10^{-3}$ & 9.30 $\times 10^{-2}$ & 0.270 &
0.358 \\
\hline
\end{tabular}
\end{center}
\end{table}
\section{Results}
In order to compare the power of our estimators, we have obtained the
mean number, eccentricity, Gaussian curvature and 
genus for a large number of Gaussian and non-Gaussian simulations
with two different angular resolutions which correspond to the LFI
100 GHz and the HFI 217 GHz channels of the Planck satellite. We have
chosen a value of $a=0.8$ for the non-Gaussian maps, that introduces
small but interesting 
differences. We have also investigated the effect of instrumental noise.
In order to have a
reasonable number of spots at any given threshold in all of our realisations
we have considered thresholds in the range $-2.5 \le \nu \le 2.5$
($-2 \le \nu \le 2.5$ for the $P$ distribution in some cases). 
In most of our cases thresholds close to $\nu=0$ do not contain very
relevant information and have not been included to speed up the
numerical calculations. 

\subsection{Noiseless case}
%
%
%
\begin{figure*}
\resizebox{\hsize}{!}{\includegraphics
{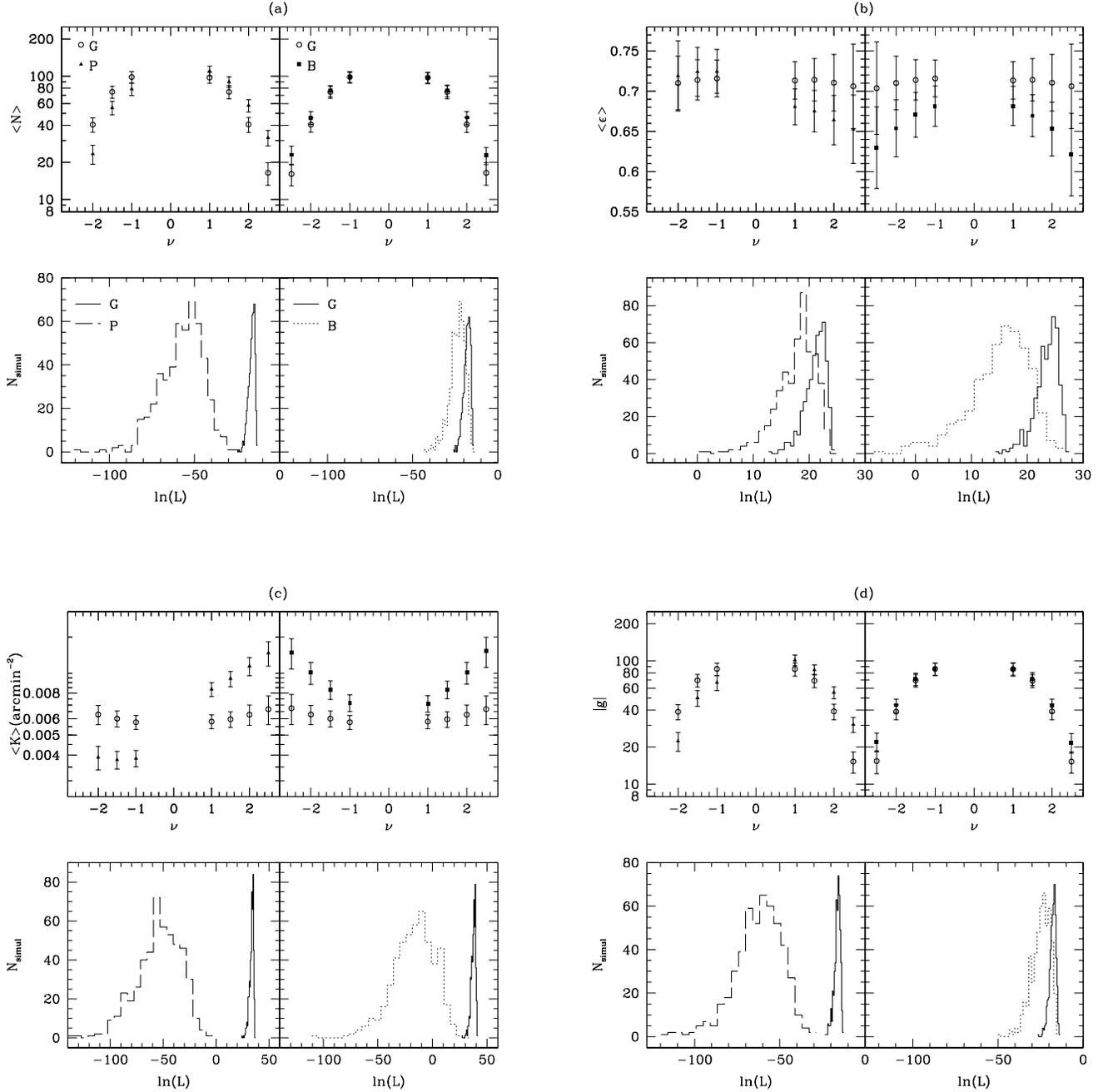}}
\caption[]{A comparison of the power of our estimators on detecting
small scale non-Gaussianity using simulated maps with the
specifications of the Planck 100 GHz channel (no noise) is shown. 
The four figures in the top-left corner (labelled (a)) correspond to
the results for the number of spots above (for positive thresholds) or
below (for negative thresholds) $\nu$. The two top figures of this set
show the mean values and dispersions obtained from the distribution of
the number of excursion sets over 500 simulations for the Gaussian
(open circles), $P$ (solid triangles) and $B$ (solid squares)
fields. The histogram of likelihood values computed from the same
simulations are plotted in the two corresponding bottom figures
for the Gaussian (solid line), $P$ (long-dashed line) and $B$
(short-dashed line) distributions.
The same results are given for the  eccentricity (top-right figures, labelled
(b)), Gaussian curvature (lower-left figures, (c)) and genus
(lower-right figures, (d)).}
\label{freq100}
\end{figure*}
First we will consider simulated maps where no noise has been
included. The top panels of Fig.~\ref{freq100}a show the
mean value and dispersion of the number of spots above (for positive
$\nu$) or below (for negative $\nu$) a threshold obtained over 500 Gaussian and
non-Gaussian realisations of size $256 \times 256$ (pixel 3$'$) and smoothed
with a Gaussian beam of FWHM=$10'$ (LFI 100 GHz channel). Open
circles, solid triangles and solid squares correspond to the Gaussian, $P$ and
$B$ fields respectively.
In order to establish a better comparison we have constructed the
likelihood function for the Gaussian and non-Gaussian maps (compared
to the Gaussian model) given by
\begin{eqnarray}
L&=&\frac{1}{\sqrt{2\pi}|M|^{1/2}}e^
{-\frac{1}{2}\left(x^s-<x^G>\right)^TM^{-1}\left(x^s-<x^G>\right)}\!, \\
M_{ij}&=&<(x_i^G-<x_i^G>)(x_j^G-<x_j^G>)> \nonumber
\end{eqnarray}
where $s$ refers to the Gaussian and non-Gaussian simulated data
and $G$ to the Gaussian ones. $M$ is computed from the Gaussian realisations.
The histogram of the likelihood values for our Gaussian and
non-Gaussian fields when using the number of spots are plotted in the
bottom panels of Fig.~\ref{freq100}a.
The mean values, dispersions and likelihood curves for the
eccentricity, Gaussian curvature and genus (in absolute value) are given in
Figs.~\ref{freq100}b,~\ref{freq100}c and~\ref{freq100}d respectively.
For negative thresholds the Gaussian curvature and eccentricity
correspond to the excursion sets below that threshold.
We can see in these figures that the $P$ distribution can be clearly
distinguished from the Gaussian model using the number, Gaussian
curvature or the genus since 
the likelihood curves do not overlap, whereas the eccentricity does
not discriminate between both distributions. In the three first cases,
the $100\%$ of the non-Gaussian simulations lie outside the $99\%$ 
confidence limit of the Gaussian case, whereas only $23.8\%$ of the $P$
simulations could be distinguished from the Gaussian model at the
$99\%$ c.l. using the eccentricity.
To establish a better comparison in the performance of our estimators,
we have calculated a factor $d$ defined as
\begin{equation}
d=\frac{m_G-m_{NG}}{\sigma_G}
\end{equation}
where $m_G$ and $m_{NG}$ correspond to the median of the
likelihood curve of the Gaussian and non-Gaussian cases respectively
and $\sigma_G$ is the dispersion of the likelihood curve for the
Gaussian simulations. 
Therefore this quantity is an indication of how well we can separate
between the Gaussian and non-Gaussian distribution with a given
estimator. Table~\ref{dfactor} gives the values of this quantity for
the different cases considered in this paper without including
instrumental noise. In the same table, the quantity $f$ indicates the
percentage of non-Gaussian simulations that lies outside the Gaussian $99\%$
c.l. for a given estimator.
We can see that for the $P$ distribution in the case of the LFI 100 GHz
channel, the highest value of $d$ (and therefore the best estimator)
corresponds to the Gaussian curvature ($d$=44.8 versus
$d$=24.7 for the genus, the next best quantity).
Regarding the $B$ model, the likelihood curves overlap for all the
estimators, except for the Gaussian curvature, that gives again the
best results ($d$=26.1, $f$=100$\%$). The other three estimators can not
discriminate between both distributions, being the eccentricity
($d$=3.9, $f$=60.6$\%$) slightly better than the genus
($d$=3.4,$f$=56.2$\%$) and the number ($d$=2.6,$f$=45.8$\%$).
\begin{center}
\begin{table*}
\caption[]{Values of $d$ and $f$ (no noise)}
\label{dfactor}
\begin{tabular}{|c|c|c|c|c|c|c|c|c|c|}
\hline
& & \multicolumn{4}{|c|}{$P$ distribution} & \multicolumn{4}{|c|}{$B$
distribution} \\
\hline
Channel & Factor & $N$ & $\varepsilon$ & $\kappa$ & $g$ & $N$ & $\varepsilon$ &
$\kappa$ & $g$ \\
\hline
100 GHz & $d$ & 21.6 & 1.6  & 44.8 & 24.7& 2.6  & 3.9  & 26.1& 3.4 \\ 
	& $f(\%)$ & 100  & 23.8 & 100  & 100 & 45.8 & 60.6 & 100 & 56.2 \\
\hline
217 GHz & $d$ & 28.5& 2.0  & 67.1& 24.1& 2.5  & 3.9  & 24.9& 2.4 \\  
	& $f(\%)$ & 100 & 23.4 & 100 & 100 & 41.6 & 58.4 & 100 & 36.0 \\
\hline
\end{tabular}
\end{table*}
\end{center}
%
%
\begin{figure*}
\resizebox{\hsize}{!}{\includegraphics
{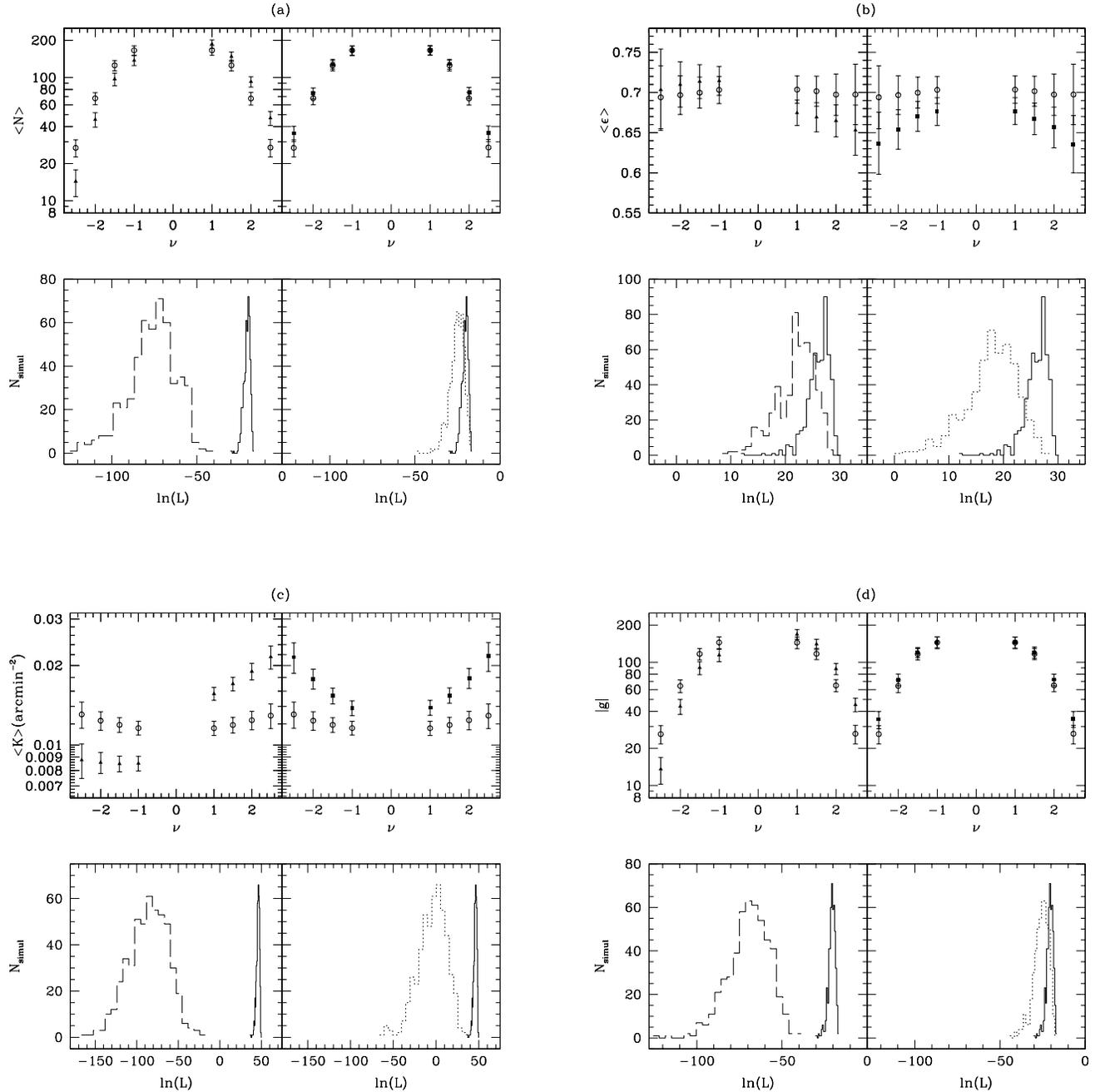}}
\caption[]{As in Fig.\ref{freq100}, but with the results corresponding to the
217 GHz channel.}
\label{freq217}
\end{figure*}

The results for the small simulated maps of the 217 GHz channel
($512\times 512$ maps,
pixel=$1'.5$, FWHM=$5'.5$) are shown in
Figs.~\ref{freq217}a (number),~\ref{freq217}b
(eccentricity),~\ref{freq217}c (curvature) and~\ref{freq217}d (genus).
As before, the $P$ distribution can be clearly distinguished from the
Gaussian one using the Gaussian curvature, genus or number of
excursion sets but not with the eccentricity. In particular the
Gaussian curvature is again the best of our statistics with $d$=67.1
and $f$=100$\%$ whereas the mean number ($d$=28.5, $f$=100$\%$) and genus
($d$=24.1, $f$=100$\%$) produce very similar results.
The $<\kappa>$ also succeeds in distinguishing between the $B$ and
Gaussian distributions ($d$=24.9,$f$=100$\%$), failing the rest of estimators
to discriminate between them.

It is interesting to point out that, in spite of the better resolution
of the 217 GHz channel, the level of success of our estimators is very
similar for both the 100 and 217 GHz channels.
This can be understood taking into account that the coherence angle
for the considered $\Lambda$-model is $\simeq 10'$.
Therefore the information about the intrinsic structure of the CMB
contained in both channels is very similar.

\subsection{The effect of instrumental noise}

\begin{figure*}
\resizebox{\hsize}{!}{\includegraphics
{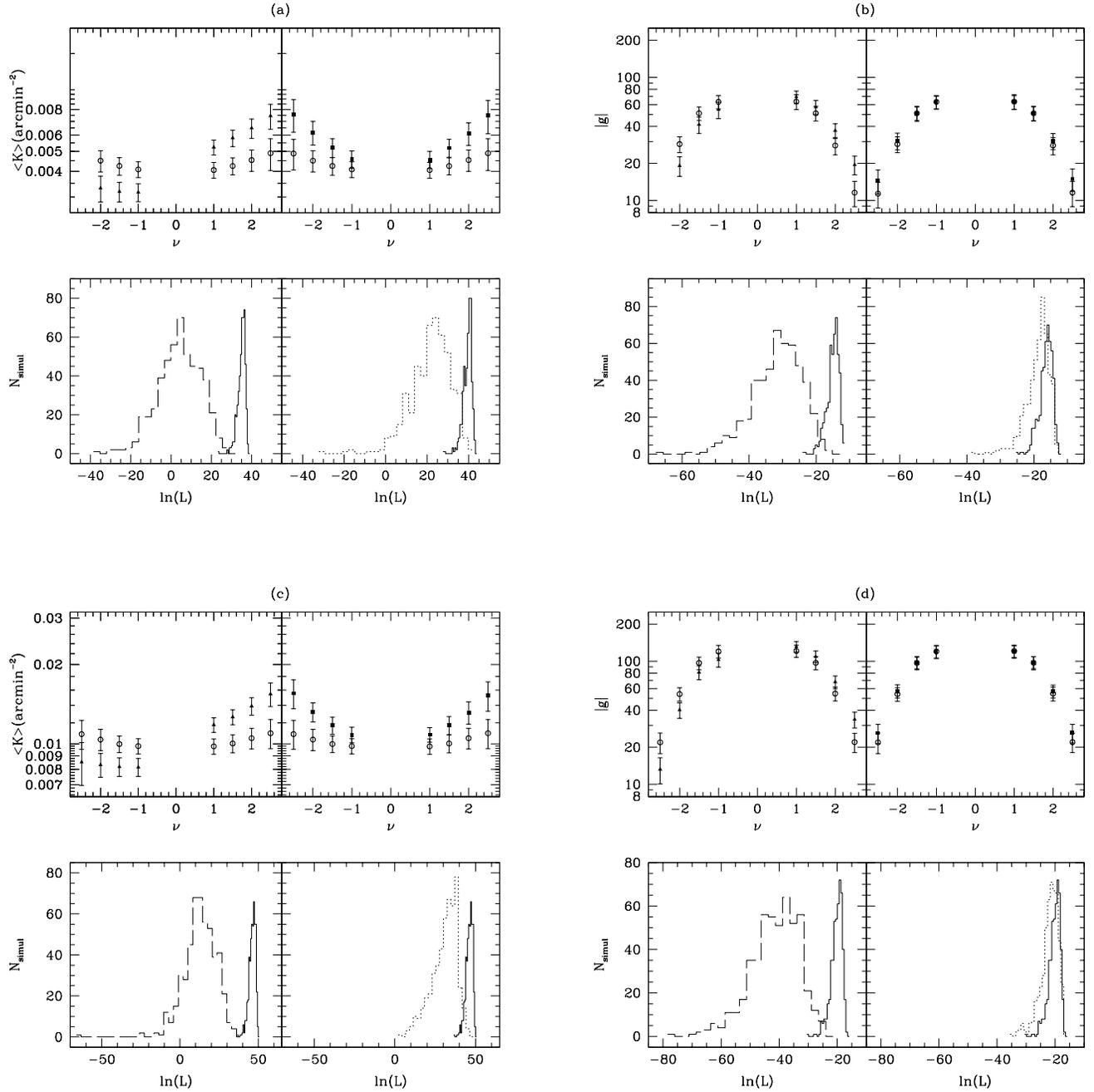}}
\caption[]{The results for the Gaussian curvature-100 GHz (four top-left
figures, (a)), genus-100 GHz (top-right figures, (b)),
Gaussian curvature-217 GHz (bottom-left figures, (c)) and genus-217 GHz
(bottom-right, (d)) are shown. Noise has been included in the simulations.}
\label{noisy}
\end{figure*}

We have also investigated the effect of instrumental noise in the
performance of our estimators. In particular, we have included white
Gaussian noise at the level predicted for the Planck satellite after
12 months of observation for the two channels that we are
considering ($\Delta T/T $ per resolution element = $4.3 \times 10^{-6}$
in both cases). For our model this corresponds to a signal to noise
ratio ($S/N$) per pixel of $S/N\simeq 2.8$ and $S/N\simeq 2.7$ for the 100 
and 217 GHz channels respectively.
In order to increase the $S/N$, the simulated signal
plus noise maps have been smoothed with a Gaussian beam with FWHM equal to the
antenna of the corresponding channel.
\begin{center}
\begin{table*}
\caption[]{Values of $d$ and $f$ (noise included)}
\label{dfactornoise}
\begin{tabular}{|c|c|c|c|c|c|c|c|c|c|}
\hline
& & \multicolumn{4}{|c|}{$P$ distribution} & \multicolumn{4}{|c|}{$B$
distribution} \\
\hline
Channel & Factor & $N$ & $\varepsilon$ & $\kappa$ & $g$ & $N$ & $\varepsilon$ &
$\kappa$ & $g$ \\
\hline
100 GHz & $d$ & 7.5 & 0.4 & 16.2 & 8.9 & 0.9 & 0.8 & 8.2 & 1.0 \\
	& $f(\%)$ & 93.6 & 5.0 & 100 & 97.6 & 12.4 & 10.0 & 89.0 & 19.8 \\
\hline
217 GHz & $d$ & 11.4 & 1.1 & 16.4 & 10.5 & 0.9 & 1.5 & 6.8 & 0.7 \\
	& $f(\%)$ & 99.6 & 7.0 & 100 & 99.2 & 10.0 & 12.0 & 93.0 & 11.4 \\
\hline
\end{tabular}
\end{table*}
\end{center}
The results are summarised in Table~\ref{dfactornoise} and
Figs.~\ref{noisy}a (Gaussian curvature, 100 GHz),~\ref{noisy}b
(genus, 100 GHz),~\ref{noisy}c (Gaussian curvature, 217 GHz) and~\ref{noisy}d
(genus, 217 GHz). Only the two best estimators are shown.
%
%
The most interesting point is that again the Gaussian curvature is the
best estimator in all the considered cases. In particular for the $P$
distribution the $100\%$ of the realisations have lied
outside the $99\%$ confidence level of the Gaussian model for both
frequency channels. The genus and the number of excursion sets give again
very similar results. They are able to discriminate between most of
the $P$ and Gaussian realisations, although the confusion level is
larger than for the Gaussian curvature. The eccentricity completely
fails to distinguish between them.

For the $B$ distribution, there is a small overlap
between both likelihood curves when using the Gaussian curvature but
still most of the realisations fall 
outside the Gaussian $99\%$ c.l. ($d$=6.8,$f$=93$\%$ for the 100 GHz channel,
$d$=8.2,$f$=89$\%$ for the 217 GHz channel). The number, genus and
eccentricity can not discriminate between the $B$ and
Gaussian distributions having values of $d \sim 1$.

\section{Conclusions}
We have performed a comparative study of the ability of three
geometrical estimators (mean 
number, eccentricity and Gaussian curvature of excursion sets above or
below a threshold) as well as the genus on
detecting small scale non-Gaussianity in the CMB. To test such
estimators, we have 
generated two kind of non-Gaussian realisations that are perturbations
of a Gaussian field. The first one ($P$ distribution) introduces a
certain level of 
skewness in the 1-pdf whereas the second one ($B$ distribution)
has a non-negligible kurtosis.
This could resemble the characteristics of some topological defect
models. The simulated maps are small patches of the sky (12.8 square degrees)
with angular resolutions which correspond approximately to two of the
Planck satellite frequency channels, the LFI 100 GHz channel
(pixel size of $3'$, FWHM=10$'$) and the
HFI 217 GHz (pixel size of $1'.5$,
FWHM=5$'$.5). We have also taken into account the expected level of
instrumental noise in these channels.

We find that the Gaussian curvature
is the best of our estimators. This is a robust conclusion, since this
quantity performs considerably better than the genus, number and eccentricity
of excursion sets for both types of non-Gaussian fields, in
presence or absence of noise and for the two considered angular
resolutions.
Regarding the number of excursion sets and the genus, these two
quantities carry very similar information, as one would expect. They
can discriminate between the $P$ and Gaussian models at a fairly high
level (although always worse than the Gaussian curvature) in all the
considered cases. However, these quantities fail to discriminate
between the $B$ and Gaussian distributions, being the level of
confusion especially high when instrumental noise is introduced.
The performance of the eccentricity is very
poor. In none of the considered cases can clearly discriminate between
the non-Gaussian and Gaussian fields, although it is 
interesting to point out that it does perform at similar or better
levels than the genus and the number of excursion sets when testing the $B$
distribution.
Therefore, the main conclusion of our work is that the mean Gaussian curvature
of excursion sets could be a very useful estimator to look for small
scale non-Gaussianity in the CMB.

\section*{Acknowledgements}
The authors warmly thank Laura Cay\'on, Jose M. Diego, Pedro Ferreira,
Aled Jones and Mike Hobson for useful discussions related to this work.
RBB acknowledges financial support from the PPARC in the form of a
research grant. EMG and JLS thank the Comisi\'on Conjunta
Hispano-Norteamericana de Cooperaci\'on Cient\'\i fica y Tecnol\'ogica
ref. 98138, Spanish DGESIC Project no. PB98-0531-c02-01, FEDER project
no. 1FD97-1769-c04-01  and the EEC project INTAS-OPEN-97-1992 for
partial financial support. RBB thanks the Instituto de F\'\i sica de
Cantabria for its hospitality during a stay in 2000.


\begin{thebibliography}{}
\bibitem[]{}Barreiro R.B., 2000, New Astron. Rev., 44, 179
\bibitem[]{} Barreiro R.B., Hobson M.P., Lasenby A.N., Banday A.J.,
G\'orski K.M. \& Hinshaw G., 2000, MNRAS, in press
\bibitem[]{} Barreiro R.B., Sanz J.L., Mart\'\i nez-Gonz\'alez E., 
Cay\'on L. \& Silk J., 1997, ApJ, 478, 1 
\bibitem[]{} Barreiro R.B., Sanz J.L., Mart\'\i nez-Gonz\'alez E., 
\& Silk J., 1998, MNRAS, 296, 693
\bibitem[]{}Bennett C.L. et al., 1997, AAS, 191, 8701; MAP homepage
http://map.gsfc.nasa.gov 
\bibitem[]{} Bouchet F.R., Peter P., Riazuelo A. \& Sakellariadou,
2000, Phys.Rev.Lett., submitted (astro-ph/0005022) 
\bibitem{} Coles P. \& Barrow J.D., 1987, MNRAS, 228, 407
\bibitem[]{} Coles P., 1988, MNRAS, 234, 509
\bibitem[]{} Contaldi C.R., 2000, Phys.Rev.Let., submitted
(astro-ph/0005115)
\bibitem[]{} de Bernardis, P. et al., 2000, Nature, 404, 955
\bibitem[]{} Diego J.M., Mart\'\i nez-Gonz\'alez E., Sanz J.L.,
Mollerach S. \& Mart\'\i nez V.J., 1999, 306, 427
\bibitem{} Ferreira P.G., Magueijo J. \& G\'orski K.M., 1998, ApJ,
503, L1
\bibitem{} Gott III J.R., Park C., Juszkiewicz R., Bies W.E., Bennett
D.P., Bouchet F.R. \& Stebbins A., 1990, ApJ, 352, 1
\bibitem[]{} Hamilton A.J.S., Gott III J.R. \& Weinberg D., 1986,
ApJ, 309, 1
\bibitem[]{}Hanany S. et al., 2000, ApJL, submitted (astro-ph/0005123)
\bibitem{} Heavens A.F., 1998, MNRAS, 299, 805
\bibitem[]{}Heavens A.F. \& Sheth R.K., 1999, MNRAS, 310, 1062
\bibitem[]{} Hobson M.P., Jones A.W. \& Lasenby A.N., 1999, MNRAS,
309, 125
\bibitem{} Hu W., Seljak U., White M. \& Zaldarriaga M. 1998,
Phys.Rev., D57, 3290
\bibitem[]{}Kogut A., Banday A.J., Bennett C.L., G\'orski K., 
Hinshaw G., Smoot G.F. \& Wright E.L., 1996, ApJ, 464, L29
\bibitem{} Jones A.W., 1998, `Application of novel analysis 
techniques to Cosmic Microwave Background Astronomy', PhD thesis,
U.Cambridge
\bibitem[]{}Jones M.E. \& Scott P.F. 1998, in J.Tr\^an Thanh V\^an,
et al. (Eds.),
Fundamental Parameters in Cosmology, Proc. of the XXXIIIrd Recontres
de Moriond (France), Editions Fronti\'eres, p.233
\bibitem[]{}Magueijo J., 2000, ApJ, 528, L57 (see also Magueijo J.,
2000, ApJ, 532, L157)
\bibitem{} Mart\'\i nez-Gonz\'alez E., Barreiro R.B., Diego J.M.,
Sanz J.L., Cay\'on L., Silk J., Mollerach S. \& Mart\'\i nez V.J.,
2000, ApL\&C, in press
\bibitem[]{}Mukherjee P., Hobson M.P. \& Lasenby A.N., 2000, MNRAS, in
press (astro-ph/0001385)
\bibitem[]{}Pando J., Valls-Gabaud D. \& Fang L., 1998,
Phys.Rev.Lett., 81, 4568
\bibitem[]{} Perivolaropoulos L., 1993, Phys.Rev., D48, 1530
\bibitem{} Pompilio M.P., Bouchet F.R., Murante G. \& Provenzale A., 1995,
ApJ, 449, 1
\bibitem{} Seljak U. \& Zaldarriaga M., 1996, ApJ, 469, 437
\bibitem[]{}Tauber J.A., 2000, ApL\&C, in press; Planck homepage 
http://astro.estec.esa.nl/Planck/ 
\bibitem[]{} Turok N., 1996, ApJ, 473, L5
\bibitem{} Vanmarcke E.H., 1983, `Random fields: Analysis and
Synthesis', MIT Press, Cambridge, MA
\end{thebibliography}
\end{document}